\documentclass[conference]{IEEEtran}
\IEEEoverridecommandlockouts
\usepackage{amsmath,amssymb,amsfonts}
\usepackage{algorithmic}
\usepackage{algorithm}
\usepackage{array}
\usepackage{textcomp}
\usepackage{stfloats}
\usepackage{url}
\usepackage{verbatim}
\usepackage{graphicx}
\usepackage{epstopdf}
\usepackage{cite}
\usepackage[subtle,tracking=normal]{savetrees}
\usepackage{subcaption}
\usepackage{xcolor}
\usepackage{caption}
\usepackage{multirow}
\usepackage[left=0.68in,right=0.673in,top=0.7in,bottom=1.013in]{geometry}
\def\BibTeX{{\rm B\kern-.05em{\sc i\kern-.025em b}\kern-.08em
    T\kern-.1667em\lower.7ex\hbox{E}\kern-.125emX}}
    
\newcommand{\removelatexerror}{\let\@latex@error\@gobble}
\makeatother

\usepackage{stackengine}

\definecolor{mintbg}{rgb}{.63,.79,.95}
\usepackage{enumitem}
\makeatletter
\g@addto@macro\normalsize{%
  \setlength\abovedisplayskip{5.5pt plus 2pt minus 1pt}%
  \setlength\belowdisplayskip{5.5pt plus 2pt minus 1pt}%
  \setlength\abovedisplayshortskip{2.5pt plus 2pt}%
  \setlength\belowdisplayshortskip{2.5pt plus 2pt}}
\makeatother
\begin{document}

%
\title{{Joint Discrete Pinching-Antenna Activation and Beamforming Design for Efficient PASS-ISAC}}
\author{\IEEEauthorblockN{Joshua Ogundiya\IEEEauthorrefmark{1}, Animesh Yadav\IEEEauthorrefmark{1} and Octavia Dobre\IEEEauthorrefmark{2}\\
\IEEEauthorblockA{\IEEEauthorrefmark{1}School of EECS, Ohio University, Athens, OH, USA\\ 
\IEEEauthorrefmark{2}Faculty of Engineering and Applied Science, Memorial University, St. John’s, NL, Canada\\
Emails: \{jo741324, yadava\}@ohio.edu, odobre@mun.ca}}}


\maketitle
\begin{abstract}
Pinching-antenna systems (PASS) offer flexible radiation points for integrated sensing and communication (ISAC), but existing designs largely assume continuous antenna placement, which may be impractical with pre-installed hardware. We investigate a discrete-activation multi-waveguide PASS-ISAC system that selects sparse transmit and receive pinching antennas (TPAs/RPAs) from predefined locations. We formulate a communication-rate maximization problem by jointly optimizing beamforming and TPA/RPA activation under transmit-power, activation, and sensing signal-to-noise ratio (SNR) constraints. To solve this nonconvex problem, we develop a low-complexity joint discrete pinching-antenna activation and beamforming optimization (JDPABO) algorithm based on alternating optimization, successive convex approximation, and reweighted $\ell_1$ minimization. Simulations show that JDPABO approaches continuous-placement performance at low sensing SNR requirements and outperforms fixed activation schemes with sparse PA activation.

\end{abstract}
\section{Introduction}
Wireless networks are evolving to support ubiquitous, high-rate, and ultra-reliable connectivity \cite{wen2024iscc}, yet line-of-sight (LoS) blockage and large-scale path loss remain key challenges \cite{liu2025passOutlook}. While reconfigurable intelligent surfaces (RIS), fluid antenna systems (FAS), and movable antennas (MA) enhance channel conditions through adaptive reconfiguration \cite{pan2020ris6g,wu2024fas6g,zhu2023ma}, their limited spatial range may be insufficient under severe blockage \cite{yang2025pinchingPrinciples}. Pinching antenna systems (PASS) address this limitation by enabling configurable radiating points along low-loss dielectric waveguides \cite{suzuki2022pinching,ouyang2025passModeling,Wijewardhana2026WiPASS,Xu2026GeneralizedPASS}. By activating antennas near users or sensing targets, PASS can strengthen LoS links, reduce path loss, and provide additional positional degrees of freedom \cite{ding2024pinchingPerspective,Yang2026PinchingAntennas}. Joint PA selection and resource management can further enhance PASS performance \cite{Zeng2026PASSRA}.

By flexibly positioning radiating points, PASS enhances communication and sensing performance, making it well-suited for integrated sensing and communication (ISAC). Recent studies have explored this potential from different perspectives \cite{zhang2025passISAC, ding2025crlbPASSISAC,mao2025multiWaveguidePASSISAC,qin2025passISACPositionPower,xu2025passPowerRadiation}. In \cite{zhang2025passISAC}, the authors proposed a two-waveguide PASS-ISAC architecture with separate waveguides for ISAC transmission and echo reception. In \cite{ding2025crlbPASSISAC}, the authors characterized the sensing capability of PASS using the Cram\'er-Rao lower bound (CRLB) and demonstrated improved sensing accuracy over conventional antennas. In \cite{mao2025multiWaveguidePASSISAC}, the authors investigated a multi-waveguide PASS-ISAC system with joint antenna-position and beamforming optimization, while the authors in \cite{qin2025passISACPositionPower} employed reinforcement learning to jointly optimize antenna positions and transmit power. Furthermore, in \cite{xu2025passPowerRadiation}, the authors developed a discrete-activation PASS framework with joint optimization of beamforming and antenna activation.

Although prior studies have demonstrated the potential of PASS for communication and ISAC, most assume continuous antenna positioning, which may be impractical due to hardware constraints and predefined activation sites. Discrete activation provides a practical alternative by selecting antennas at fixed locations. However, existing studies \cite{xu2025passPowerRadiation,Wang2025AntennaActivation,Zhao2026WaveguideDivision} primarily address communication systems, leaving multi-waveguide PASS-ISAC largely unexplored. To bridge this gap, we propose a discrete-activation framework for multi-waveguide PASS-ISAC, with the following contributions:
\begin{itemize}
  \item We consider a multi-waveguide PASS-ISAC system with discrete TPA/RPA activation and characterize communication and sensing performance through achievable rate and sensing SNR.

  \item We formulate a joint beamforming and sparse TPA/RPA activation problem and develop a low-complexity joint discrete pinching-antenna activation and beamforming optimization (JDPABO) algorithm using reweighted $\ell_1$ minimization and successive convex approximation (SCA).

  \item We demonstrate via simulation that JDPABO effectively balances communication and sensing while requiring only a limited number of active TPAs/RPAs.
\end{itemize}

The remainder of this paper is organized as follows. Section II presents the PASS-ISAC system model and problem formulation. Section III develops the proposed solution, Section IV presents the simulation results, and Section V concludes the paper.

\emph{Notations:} Boldface lowercase and uppercase letters denote vectors and matrices, respectively. $(\cdot)^T$ and $(\cdot)^H$ denote transpose and Hermitian transpose. $\|\cdot\|_0$, $\|\cdot\|_1$, and $\|\cdot\|$ denote the $\ell_0$-, $\ell_1$-, and $\ell_2$-norms, respectively. $\mathrm{diag}(\cdot)$ and $\mathrm{blkdiag}(\cdot)$ denote diagonal and block-diagonal operators. $\mathbb{C}$ and $\mathbb{R}$ denote the complex and real number sets, while $\Re(\cdot)$, $\Im(\cdot)$, and $\mathbb{E}[\cdot]$ denote the real part, imaginary part, and expectation, respectively.

\section{System Model}
\begin{figure}[t]
    \centering
    \includegraphics[width=\linewidth]{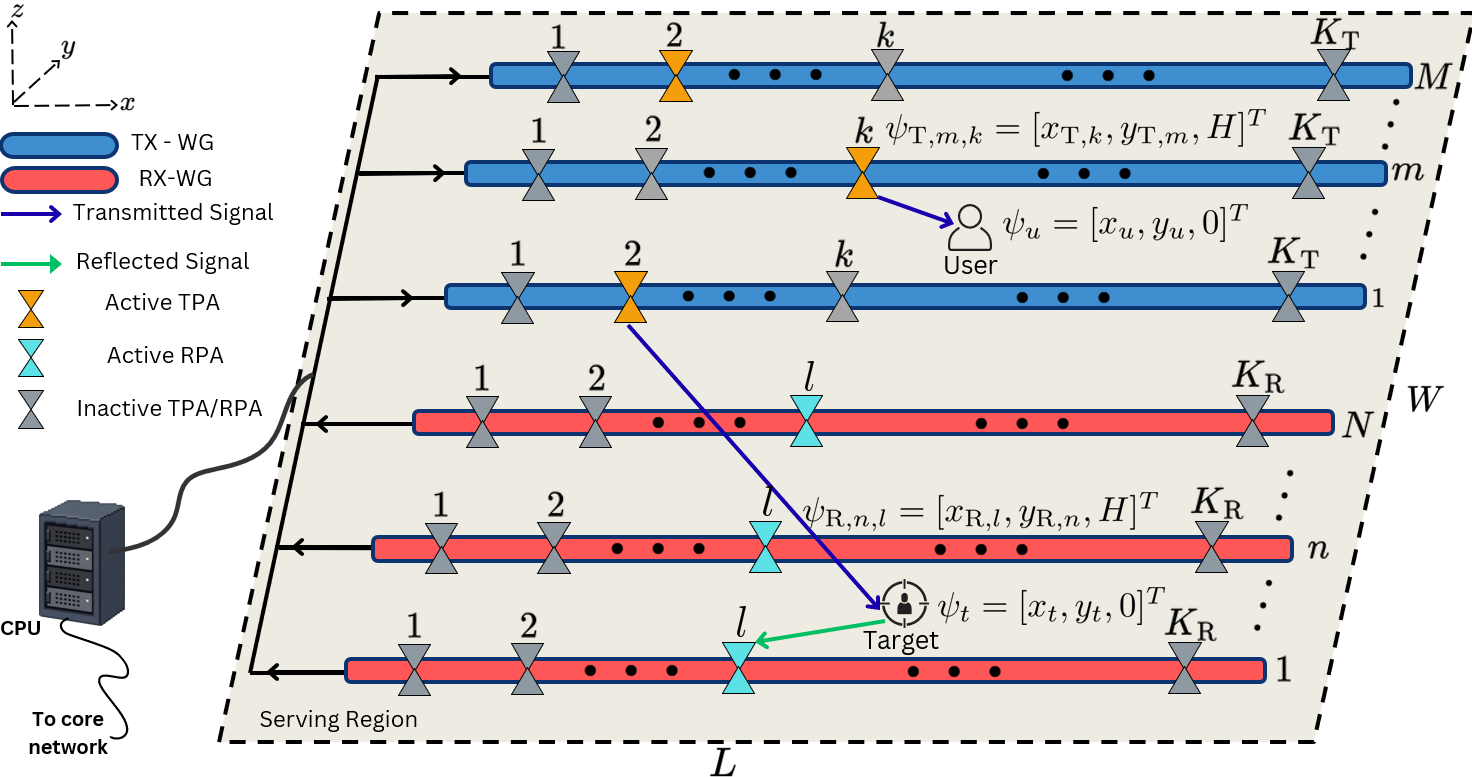}
    \caption{Multi-waveguide pinching-antenna based ISAC system.}
    \label{fig:system_model}
\end{figure}

We consider a multi-waveguide PASS-ISAC system consisting of a dual-function central processing unit (CPU) that communicates with a single-antenna user while simultaneously sensing a point target, as illustrated in Fig.~\ref{fig:system_model}. The CPU is connected to $M$ transmit and $N$ receive dielectric waveguides, indexed by $\mathcal{M}=\{1,\ldots,M\}$ and $\mathcal{N}=\{1,\ldots,N\}$, respectively. Each transmit (receive) waveguide contains $K_{\text{T}}$ TPAs ($K_{\text{R}}$ RPAs) pre-mounted at discrete candidate locations, indexed by $\mathcal{K}_{\text{T}}=\{1,\ldots,K_{\text{T}}\}$ ($\mathcal{K}_{\text{R}}=\{1,\ldots,K_{\text{R}}\}$). At each time instant, only a subset of these TPAs and RPAs is activated, resulting in sparse transmit and receive activation patterns. We adopt a three-dimensional (3D) Cartesian coordinate system, wherein the waveguides are deployed at height $H$ over a rectangular serving area of length $L$ and width $W$ in the $x$-$y$ plane, with the midpoint of each waveguide centered at $x=0$ and its $y$-coordinate fixed by the deployment layout.

Accordingly, the 3D coordinates of the $k$th TPA on the $m$th transmit waveguide and the $l$th RPA on the $n$th receive waveguide are given by $\boldsymbol{\psi}_{\text{T},m,k}=[x_{\text{T},k},y_{\text{T},m},H]^T$ and $\boldsymbol{\psi}_{\text{R},n,l}=[x_{\text{R},l},y_{\text{R},n},H]^T$, respectively. The candidate PA sites are uniformly spaced along each waveguide, with $\Delta x_{\text{T}}=L/(K_{\text{T}}-1)$ and $\Delta x_{\text{R}}=L/(K_{\text{R}}-1)$. Thus, the discrete positions are $x_{\text{T},k}=-L/2+(k-1)\Delta x_{\text{T}}$, $k\in\mathcal{K}_{\text{T}}$, and $x_{{\text{R}},l}=-L/2+(l-1)\Delta x_{\text{R}}$, $l\in\mathcal{K}_{\text{R}}$, forming the candidate TPA and RPA sets $\mathcal{X}_{\text{T}}=\{x_{\text{T},1},\ldots,x_{\text{T},K_{\text{T}}}\}$ and $\mathcal{X}_{\text{R}}=\{x_{\text{R},1},\ldots,x_{\text{R},K_{\text{R}}}\}$, respectively. The user and target are located on the ground at $\boldsymbol{\psi}_u=[x_u,y_u,0]^T$ and $\boldsymbol{\psi}_t=[x_t,y_t,0]^T$, respectively.

\subsection{Channel Model}
Due to near-field propagation, we adopt the spherical-wave channel model \cite{zhang2022beamfocusing}. Accordingly, the free-space channel between the $k$th TPA of the $m$th waveguide and user $u$ is given as
\begin{equation}
h_{u,m,k}
=
\frac{
\sqrt{\eta}\,
\exp\!\left(
- j\frac{2\pi}{\lambda}\,
\left\| \boldsymbol{\psi}_u - \boldsymbol{\psi}_{\text{T},m,k} \right\|
\right)
}{
\left\| \boldsymbol{\psi}_u - \boldsymbol{\psi}_{\text{T},m,k} \right\|
},
\end{equation}
where $\eta = \frac{\lambda^2}{(4\pi)^2}$ with
$\lambda$ is the free-space wavelength and
$\boldsymbol{\psi}_u$ and $\boldsymbol{\psi}_{\text{T},m,k}$ denote the 3D coordinates of the user and the $k$th TPA on the $m$th waveguide, respectively.

The channel between the feeding point of the $m$th transmit waveguide and its $k$th TPA is given as
\begin{equation}
g_{m,k}
=
\alpha_{m,k}\,
\exp\!\left(
- j\frac{2\pi}{\lambda_g}\,
\left\| \boldsymbol{\psi}_{\text{T},m}^{0} - \boldsymbol{\psi}_{\text{T},m,k} \right\|
\right),
\end{equation}
where $\alpha_{m,k}$ is the coupling coefficient between the $m$th transmit waveguide and the $k$th TPA,
$\lambda_g$ is the guided wavelength inside the dielectric waveguide,
$\boldsymbol{\psi}_{\text{T},m}^{0} = \left[-\frac{L}{2},\, y_{\text{T},m},\, H\right]^T$ is the feed-point location of the $m$th waveguide,
and $\|\boldsymbol{\psi}_{\text{T},m}^{0} - \boldsymbol{\psi}_{\text{T},m,k}\|$ is the waveguide propagation distance to the PA.

Similarly, the free-space channel between the target $t$ and the $l$th RPA on the $n$th receive waveguide is given as
\begin{equation}
f_{t,n,l}
=
\frac{
\sqrt{\eta}\,
\exp\!\left(
- j\frac{2\pi}{\lambda}\,
\left\| \boldsymbol{\psi}_t - \boldsymbol{\psi}_{\text{R},n,l} \right\|
\right)
}{
\left\| \boldsymbol{\psi}_t - \boldsymbol{\psi}_{\text{R},n,l} \right\|
},
\end{equation}
where $\boldsymbol{\psi}_{\text{R},n,l}$ denotes the 3D coordinate of the $l$th RPA on the $n$th receive waveguide, and $\|\boldsymbol{\psi}_t-\boldsymbol{\psi}_{\text{R},n,l}\|$ is the propagation distance between the target and the RPA.

The channel from the feeding point of the $n$th receive waveguide to its $l$th RPA is given by
\begin{equation}
g_{n,l}
=
\alpha_{n,l}\,
\exp\!\left(
- j\frac{2\pi}{\lambda_g}\,
\left\| \boldsymbol{\psi}_{\text{R},n}^{0} - \boldsymbol{\psi}_{\text{R},n,l} \right\|
\right),
\end{equation}
where $\alpha_{n,l}$ is the coupling coefficient between the $n$th receive waveguide and the $l$th RPA, $\boldsymbol{\psi}_{\text{R},n}^{0} = \left[-\frac{L}{2},\, y_{\text{R},n},\, H\right]^T$ is the feeding location of the $n$th receive waveguide, and $\|\boldsymbol{\psi}_{\text{R},n}^{0}-\boldsymbol{\psi}_{\text{R},n,l}\|$ denotes the waveguide propagation distance to the RPA.

Similarly, the free-space channel from the $k$th TPA on the $m$th waveguide to the target is given by
\begin{equation}
h_{t,m,k}
=
\frac{
\sqrt{\eta}\,
\exp\!\left(
- j\frac{2\pi}{\lambda}\,
\left\| \boldsymbol{\psi}_t - \boldsymbol{\psi}_{\text{T},m,k} \right\|
\right)
}{
\left\| \boldsymbol{\psi}_t - \boldsymbol{\psi}_{\text{T},m,k} \right\|
}.
\end{equation}


\subsection{Communication Model}
Let $s$ denote the unit energy (i.e., $\mathbb{E}[|s|^2]=1$) downlink data symbol corresponding to the user,  and let $w_m\in\mathbb{C}$ denote the beamforming weight for the $m$th transmit waveguide. The activation state of the $k$th TPA on waveguide $m$ is represented by $a_{m,k}\in\{0,1\}$, where $a_{m,k}=1$ indicates that the TPA is active and $a_{m,k}=0$ otherwise. Accordingly, the TPA activation vector is defined as $\mathbf{a}_m=[a_{m,1},a_{m,2},\ldots,a_{m,K_{\text{T}}}]^{T}\in\{0,1\}^{K_\text{T}}$, subject to the following constraint:
\begin{equation}
\|\mathbf{a}_m\|_0 \le  N_{\mathrm{act},\text{T}},\quad \forall m=1,\ldots,M,
\end{equation}
where $N_{\mathrm{act},\text{T}} \ll K_{\text{T}}$ is the maximum number of simultaneously active TPAs on a transmit waveguide.

The signal received at user $u$ is expressed as
\begin{equation}
y_u = \sqrt{P_{\max}}\ \mathbf{h}_u^{T}\mathbf{A}\mathbf{G}_t\mathbf{w}\, s + n_u,
\end{equation}
where $P_{\max}$ denotes the maximum transmit power budget,  $\mathbf{h}_u \triangleq 
\big[\mathbf{h}_{u,1}^{T},\,\mathbf{h}_{u,2}^{T},\,\ldots,\,\mathbf{h}_{u,M}^{T}\big]^{T}
\in \mathbb{C}^{MK_{\text{T}}\times 1}$ denotes the aggregate channel vector, with $\mathbf{h}_{u,m} = [h_{u,m,1},\,h_{u,m,2},\,\ldots,\,h_{u,m,K_{\text{T}}}]^{T}\in \mathbb{C}^{K_{\text{T}}\times 1}
$ representing the free-space channel vector between the TPAs on the $m$th waveguide and the user. The corresponding in-waveguide channels are collected in $\mathbf{G}_t \triangleq 
\mathrm{blkdiag}\!\left(\mathbf{g}_1,\,\mathbf{g}_2,\,\ldots,\,\mathbf{g}_M\right)
\in \mathbb{C}^{MK_{\text{T}}\times M}$, where $\mathbf{g}_{m}=[g_{m,1},\,g_{m,2},\,\ldots,\,g_{m,K_{\text{T}}}]^{T}
\in \mathbb{C}^{K_{\text{T}}\times 1}$. The TPA activation is represented by the diagonal selection matrix $\mathbf{A} \triangleq \mathrm{diag}(\mathbf{a}) \in \mathbb{R}^{MK_{\text{T}}\times MK_{\text{T}}}$, where 
$\mathbf{a} \triangleq \big[\mathbf{a}_{1}^{T},\,\mathbf{a}_{2}^{T},\ldots,\mathbf{a}_{M}^{T}\big]^{T}
\in \{0,1\}^{MK_{\text{T}}\times 1}$.
$\mathbf{w} \triangleq [w_1,\,w_2,\,\ldots,\,w_M]^{T}\in\mathbb{C}^{M\times 1}$ denotes the waveguide beamforming vector, and
$n_u \sim \mathcal{CN}(0,\sigma_u^{2})$ is the additive white Gaussian noise (AWGN).



\subsection{Sensing Model}
Similar to the transmit side, we introduce a binary variable $b_{n,l}\in\{0,1\}$ to represent the activation state of the $l$th RPA on the $n$th receive waveguide. For the $n$th receive waveguide, the receive activation vector is defined as $\mathbf{b}_n=[b_{n,1},b_{n,2},\ldots,b_{n,K_{\text{R}}}]^T$. Stacking all receive activation vectors gives $\mathbf{b}=[\mathbf{b}_1^T,\mathbf{b}_2^T,\ldots,\mathbf{b}_N^T]^T$, and the corresponding selection matrix is $\mathbf{B}=\mathrm{diag}(\mathbf{b})$.

For target sensing, we consider the echo signal reflected by a point-like target and collected by the active RPAs. The received echo vector is modeled as
\begin{equation}
\mathbf{y}_t
= 
\beta
\left(\sqrt{P_{\max}}\mathbf{h}_t^{T}\mathbf{A}\mathbf{G}_t\mathbf{w}\right)
\mathbf{B}\mathbf{G}_r\mathbf{f}_t s
+
\mathbf{z}_r,
\end{equation}
where $\beta\in \mathbb{C}$ denotes the complex reflection coefficient of the target, $\mathbf{z}_r\sim\mathcal{CN}(\mathbf{0},\sigma_r^2\mathbf{I})$ is the AWGN vector, and $\mathbf{w}$ is the transmit beamforming vector.

Specifically, $\mathbf{h}_t=[\mathbf{h}_{t,1}^{T},\ldots,\mathbf{h}_{t,M}^{T}]^{T}$ and $\mathbf{f}_t=[\mathbf{f}_{t,1}^{T},\ldots,\mathbf{f}_{t,N}^{T}]^{T}$, where $\mathbf{h}_{t,m}=[h_{t,m,1},\ldots,h_{t,m,K_{\text{T}}}]^{T}$ and $\mathbf{f}_{t,n}=[f_{t,n,1},\ldots,f_{t,n,K_{\text{R}}}]^{T}$ collect the corresponding free-space channel coefficients defined in the channel model. Similarly, $\mathbf{G}_r=\mathrm{diag}(\mathbf{g}_r)$, where $\mathbf{g}_r$ collects the receive-side in-waveguide channel coefficients. Let $\mathbf{u}\in\mathbb{C}^{NK_{\text{R}}\times 1}$ denote the receive beamforming vector. The post-combining sensing signal is
\begin{equation}
y_e
=
\mathbf{u}^{H}\mathbf{y}_t
=
\beta
\left(\sqrt{P_{\max}}\mathbf{h}_t^{T}\mathbf{A}\mathbf{G}_t\mathbf{w}\right)
\left(\mathbf{u}^{H}\mathbf{B}\mathbf{G}_r\mathbf{f}_t\right)s
+
\mathbf{u}^{H}\mathbf{z}_r.
\end{equation}

The target detection task can be viewed as a binary hypothesis test between the absence and presence of the target. Since the detection performance improves with the post-combining echo strength relative to the noise power, we
characterize the sensing quality using the sensing SNR. Accordingly, the sensing SNR is given by
\begin{equation}
\gamma_s
=
\frac{P_{\max}
|\beta|^{2}
|\mathbf{h}_t^{T}\mathbf{A}\mathbf{G}_t\mathbf{w}|^{2}
|\mathbf{u}^{H}\mathbf{B}\mathbf{G}_r\mathbf{f}_t|^{2}
}{
\sigma_r^{2}\|\mathbf{u}\|^{2}
}.
\end{equation}
For fixed $\mathbf{A}$, $\mathbf{B}$, and $\mathbf{w}$, maximizing $\gamma_s$ with respect to $\mathbf{u}$ is equivalent to maximizing $|\mathbf{u}^{H}\mathbf{B}\mathbf{G}_r\mathbf{f}_t|^2/\|\mathbf{u}\|^2$. According to the Cauchy--Schwarz inequality, this is achieved by choosing $\mathbf{u}$ in the direction of $\mathbf{B}\mathbf{G}_r\mathbf{f}_t$. Therefore, the optimal receive beamformer is the matched beamformer
\begin{equation}
\mathbf{u}^{\star}
=
\frac{\mathbf{B}\mathbf{G}_r\mathbf{f}_t}
{\|\mathbf{B}\mathbf{G}_r\mathbf{f}_t\|}.
\end{equation}
Substituting $\mathbf{u}^{\star}$ into $\gamma_s$ gives
\begin{equation}
\gamma_s
=
\frac{P_{\max}
|\beta|^{2}
|\mathbf{h}_t^{T}\mathbf{A}\mathbf{G}_t\mathbf{w}|^{2}
\|\mathbf{B}\mathbf{G}_r\mathbf{f}_t\|^{2}
}{
\sigma_r^{2}
}.
\end{equation}

\subsection{Problem Formulation}
With the aim of maximizing the communication rate, we jointly optimize the transmit beamforming vector and sparse TPA/RPA activation patterns subject to transmit-power, sparsity, and sensing constraints. Defining the user and target SNRs as $\gamma_u \triangleq P_{\mathrm{max}}/\sigma_u^2$ and $\gamma_r \triangleq P_{\mathrm{max}}/\sigma_r^2$, respectively, the optimization problem is formulated as
{\allowdisplaybreaks 
\begin{subequations}\label{prob:P1}
    \begin{align}
        \underset{\mathbf{w},\,\mathbf{a},\,\mathbf{b}}{\max} \quad 
        & \log_2\!\left(1+\gamma_u \left|\mathbf{h}_u^{T}\mathbf{A}\mathbf{G}_t\mathbf{w}\right|^2 \right)
        \label{eq:P1_obj}\\
        \mbox{s.t.} \quad
        & \sum_{m=1}^{M}|w_m|^2 \leq 1, \label{eq:P1_c1}\\
        & \|\mathbf{a}_m\|_0 \leq N_{\mathrm{act},\text{T}}, \quad \forall m\in\mathcal{M}, \label{eq:P1_c2}\\
        & \|\mathbf{b}_n\|_0 \leq N_{\mathrm{act},\text{R}}, \quad \forall n\in\mathcal{N}, \label{eq:P1_c3}\\
        & \gamma_r |\beta|^2 \left|\mathbf{h}_t^{T}\mathbf{A}\mathbf{G}_t\mathbf{w}\right|^2 \left\|\mathbf{B}\mathbf{G}_r\mathbf{f}_t\right\|^2 \geq \gamma_{s,\min},    \label{eq:P1_c4}\\
        & a_{m,k}\in\{0,1\}, \quad \forall m\in\mathcal{M},\; k\in\mathcal{K}_{\text{T}}, \label{eq:P1_c5}\\
        & b_{n,l}\in\{0,1\}, \quad \forall n\in\mathcal{N},\; l\in\mathcal{K}_{\text{R}}, \label{eq:P1_c6} \qquad \qquad
    \end{align}
\end{subequations}}
where $\gamma_{s,\min}$ denotes the minimum required sensing SNR. Constraint \eqref{eq:P1_c1} limits the total normalized transmit power across all transmit waveguides to unity. Constraints \eqref{eq:P1_c2} and \eqref{eq:P1_c3} limit the numbers of active TPAs and RPAs on each transmit and receive waveguide, respectively. Constraint \eqref{eq:P1_c4} ensures that the sensing SNR satisfies the minimum sensing requirement.

The formulated optimization problem is highly nonconvex and difficult to solve. In particular, the $\ell_0$-norm sparsity constraints in \eqref{eq:P1_c2} and \eqref{eq:P1_c3} lead to a combinatorial antenna-selection problem, which is NP-hard in general. Moreover, the communication objective \eqref{eq:P1_obj} and sensing constraint \eqref{eq:P1_c4} are nonconvex and coupled through the beamforming and antenna-activation variables. Therefore, we develop a computationally tractable approach to obtain a suboptimal solution.

\section{Proposed Algorithm}
We develop an algorithm using the alternating optimization (AO) framework to effectively address the nonconvex issue presented in \eqref{prob:P1}. This approach breaks down the initial problem into two distinct subproblems. First, the transmit beamforming vector $\mathbf{w}$ is optimized for fixed TPA and RPA activation patterns. Then, the TPA and RPA activation variables $\mathbf{a}$ and $\mathbf{b}$ are optimized for the obtained beamforming vector. These two subproblems are solved alternately until convergence.

\subsection{Subproblem 1}
In subproblem 1, we optimize the beamforming vector $\mathbf{w}$ for fixed binary matrices $\mathbf{A}$ and $\mathbf{B}$. Accordingly, the optimization problem \eqref{prob:P1} reduces to the following subproblem:
 \begin{subequations}\label{prob:P2}
  \begin{align}
  \max_{\mathbf{w}} \quad
  & \log_2\!\left(
  1+\gamma_u\left|\mathbf{h}_u^{T}\mathbf{A}\mathbf{G}_t\mathbf{w}\right|^2
\right) \label{eq:P2_obj}\\
\text{s.t.}\quad
& \eqref{eq:P1_c1}\text{ and } \eqref{eq:P1_c4}.\label{eq:P2_c2}
\end{align}
\end{subequations}
Problem~\eqref{prob:P2} remains nonconvex because the communication objective maximizes a convex quadratic function of $\mathbf{w}$, while the sensing constraint imposes a lower bound on a convex quadratic function, resulting in a nonconvex feasible set.
To this end, we introduce two slack variables $\rho_u \geq 0$ and $\rho_t\geq 0$ to deal with \eqref{eq:P2_obj} and \eqref{eq:P1_c4}, respectively. Consequently, the objective function equivalently becomes $\log_2(1+\rho_u)$ along with the following new constraint:
\begin{equation}
\left|\mathbf{h}_u^{T}\mathbf{A}\mathbf{G}_t\mathbf{w}\right|^2
  \ge \frac{\rho_u}{\gamma_u}, \label{eq:P2slack_user}
\end{equation}
and \eqref{eq:P1_c4} can be equivalently replaced with the following constraints:
\begin{subequations}
\begin{align} 
  &|\beta|^{2}\rho_t\|\mathbf{B}\mathbf{G}_r\mathbf{f}_{t}\|^{2}
  \ge \gamma_{s,\min}, \label{eq:P2slack_sensing}\\
  &\left|\mathbf{h}_t^{T}\mathbf{A}\mathbf{G}_t\mathbf{w}\right|^2
  \ge \frac{\rho_t}{\gamma_r}. \label{eq:P2slack_target}
  \end{align}
\end{subequations}
Constraints \eqref{eq:P2slack_user} and \eqref{eq:P2slack_target} remain
nonconvex, as they impose lower bounds on convex quadratic functions.
To obtain tractable convex constraints, we employ successive convex
approximation (SCA) based on first-order Taylor expansions \cite{Marks1978}.
To this end, we define the real and imaginary components of the
communication and sensing beamforming gains as
\begin{subequations}
\begin{align}
p_u &\triangleq \Re\{\mathbf{h}_u^{T}\mathbf{A}\mathbf{G}_t\mathbf{w}\},\quad
q_u \triangleq \Im\{\mathbf{h}_u^{T}\mathbf{A}\mathbf{G}_t\mathbf{w}\}, \label{eq:pu_qu}\\
p_t &\triangleq \Re\{\mathbf{h}_t^{T}\mathbf{A}\mathbf{G}_t\mathbf{w}\},\quad
q_t \triangleq \Im\{\mathbf{h}_t^{T}\mathbf{A}\mathbf{G}_t\mathbf{w}\}, \label{eq:pt_qt}
\end{align}
\end{subequations}
such that $p_u^2+q_u^2 \geq \frac{\rho_u}{\gamma_u}$ and $p_t^2+q_t^2 \geq \frac{\rho_t}{\gamma_r}$.

At the $j$th SCA iteration, the convex quadratic terms are replaced by
their first-order Taylor lower bounds around
$(p_u^{(j)},q_u^{(j)})$ and $(p_t^{(j)},q_t^{(j)})$, respectively.
Accordingly, \eqref{eq:P2slack_user} and \eqref{eq:P2slack_target} are
approximated by the following affine constraints:
\begin{subequations}
  \begin{align}
  \big(p_u^{(j)}\big)^2
  &+2p_u^{(j)}(p_u-p_u^{(j)})
  +\big(q_u^{(j)}\big)^2 
  +2q_u^{(j)}(q_u-q_u^{(j)})
  \ge \frac{\rho_u}{\gamma_u}, \label{eq:P2sca_user}\\
  \big(p_t^{(j)}\big)^2
  &+2p_t^{(j)}(p_t-p_t^{(j)})
  +\big(q_t^{(j)}\big)^2+2q_t^{(j)}(q_t-q_t^{(j)})
  \ge \frac{\rho_t}{\gamma_r}. \label{eq:P2sca_target}
  \end{align}
  \end{subequations}
Accordingly, the relaxed problem for finding the beamforming vector $\mathbf{w}$ at the $j$th SCA iteration is given by 
\begin{subequations}\label{prob:P2_sca}
    \begin{align}
    \max_{\mathbf{w},\,\rho_u,\,\rho_t,\,p_u,\,q_u,\,p_t,\,q_t} \quad
    & \log_2(1+\rho_u) \\
    \text{s.t.}\quad
    & \eqref{eq:P1_c1},\ \eqref{eq:P2slack_sensing},\ \eqref{eq:pu_qu},\ \eqref{eq:pt_qt}, \ \eqref{eq:P2sca_user},\ \eqref{eq:P2sca_target}.\quad
    \end{align}
\end{subequations}
Since \eqref{prob:P2_sca} is a convex problem, any standard convex optimization solver can be invoked to solve it. The resulting beamforming vector $\mathbf{w}$ is then used in the antenna activation update.

\subsection{Subproblem 2}
In subproblem 2, we optimize the TPA and RPA activation patterns for the fixed transmit beamforming vector $\mathbf{w}$. Accordingly, the optimization problem \eqref{prob:P1} reduces to the following subproblem:
\begin{subequations}\label{prob:P3}
    \begin{align}
        \max_{\mathbf{a},\,\mathbf{b}} \quad
        & \log_2\!\left(1+\gamma_u\left|\mathbf{h}_u^{T}\mathbf{A}\mathbf{G}_t\mathbf{w}\right|^2
        \right)\label{eq:P3_obj} \\
        \text{s.t.}\quad
        & \eqref{eq:P1_c2},\ \eqref{eq:P1_c3},\
        \eqref{eq:P1_c4},\ \eqref{eq:P1_c5},\ \text{and }  \eqref{eq:P1_c6}.
    \end{align}
\end{subequations}
Due to the nonconvex objective \eqref{eq:P3_obj}, sensing constraint \eqref{eq:P1_c4}, sparsity constraints \eqref{eq:P1_c2}--\eqref{eq:P1_c3}, and binary activation variables, problem~\eqref{prob:P3} remains challenging to solve. Following Subproblem~1, we introduce slack variables $\rho_u\geq0$, $\rho_t\geq0$, and $\zeta_t\geq0$ to reformulate the communication and sensing terms. Accordingly, the objective becomes $\log_2(1+\rho_u)$ subject to the following constraint:
\begin{equation}\label{eq:P3_equi_relaxed_user}
     \left|\mathbf{h}_u^{T}\mathbf{A}\mathbf{G}_t\mathbf{w}\right|^2
        \ge \frac{\rho_u}{\gamma_u}, 
\end{equation}
and \eqref{eq:P1_c4} can be equivalently replaced with the following constraints:
\begin{subequations}\label{eq:P3_equi_constraints}
    \begin{align}
        & \rho_t \zeta_t \ge \gamma_{s,\min}, \label{eq:P3_equi_bilinear} \\
        & \left|\mathbf{h}_{t}^{T}\mathbf{A}\mathbf{G}_t\mathbf{w}\right|^{2}
        \ge \frac{\rho_t}{\gamma_r}, \label{eq:P3_equi_target}\\
        & \|\mathbf{B}\mathbf{G}_r\mathbf{f}_{t}\|^{2}
        \ge \frac{\zeta_t}{|\beta|^{2}}. \label{eq:P3_equi_receive}
    \end{align}
\end{subequations}

Observe that constraints \eqref{eq:P3_equi_relaxed_user} and \eqref{eq:P3_equi_constraints} remain nonconvex. To address this, we employ SCA as in Subproblem~1. In particular, we express the bilinear term in \eqref{eq:P3_equi_bilinear} using the difference-of-convex representation as $\rho_t \zeta_t=\frac{1}{4}\left((\rho_t+\zeta_t)^2-(\rho_t-\zeta_t)^2\right)$, yielding the following constraint:
\begin{equation}
(\rho_t+\zeta_t)^2-(\rho_t-\zeta_t)^2 \ge 4\gamma_{s,\min}.
\label{eq:P3_dc_bilinear}
\end{equation}
The constraint in \eqref{eq:P3_dc_bilinear} is still nonconvex. Therefore, we replace the convex term $f(\rho_t,\zeta_t) = (\rho_t+\zeta_t)^2$ by its first-order Taylor approximation around the point$(\rho_t^{(j)}, \zeta_t^{(j)})$ as 
\begin{equation}
\begin{aligned}
f^{(j)}(\rho_t,\zeta_t)
&=
\big(\rho_t^{(j)}+\zeta_t^{(j)}\big)^2 \\
&+2\big(\rho_t^{(j)}+\zeta_t^{(j)}\big)
\Big[(\rho_t+\zeta_t)-\big(\rho_t^{(j)}+\zeta_t^{(j)}\big)\Big].
\end{aligned}
\label{eq:P3_linear_xy}
\end{equation}
Hence, the convex approximation of \eqref{eq:P3_dc_bilinear} at the $j$th SCA iteration is given by
\begin{equation}
(\rho_t-\zeta_t)^2-f^{(j)}(\rho_t,\zeta_t)+4\gamma_{s,\min}\le 0.
\label{eq:P3_bilinear_sca}
\end{equation}
To deal with \eqref{eq:P3_equi_relaxed_user}, \eqref{eq:P3_equi_target}, and \eqref{eq:P3_equi_receive}, similar to the beamforming subproblem, we define
\begin{subequations}
\begin{align}
p_u &= \Re\{\mathbf{h}_u^{T}\mathbf{A}\mathbf{G}_t\mathbf{w}\},\quad
q_u = \Im\{\mathbf{h}_u^{T}\mathbf{A}\mathbf{G}_t\mathbf{w}\}, \label{eq:P3_pu_qu}\\
p_t &= \Re\{\mathbf{h}_t^{T}\mathbf{A}\mathbf{G}_t\mathbf{w}\},\quad
q_t = \Im\{\mathbf{h}_t^{T}\mathbf{A}\mathbf{G}_t\mathbf{w}\}, \label{eq:P3_pt_qt}\\
\mathbf{p}_r &= \Re\{\mathbf{B}\mathbf{G}_r\mathbf{f}_t\},\quad
\mathbf{q}_r = \Im\{\mathbf{B}\mathbf{G}_r\mathbf{f}_t\}, \label{eq:P3_pr_qr}
\end{align}
\end{subequations}
such that
$p_u^2+q_u^2 \ge \frac{\rho_u}{\gamma_u}$, $p_t^2+q_t^2 \ge \frac{\rho_t}{\gamma_r}$ and $\|\mathbf{p}_r\|^2+\|\mathbf{q}_r\|^2 \ge \frac{\zeta_t}{|\beta|^2}$. Then, we approximate \eqref{eq:P3_equi_relaxed_user}, \eqref{eq:P3_equi_target}, and \eqref{eq:P3_equi_receive} using their first-order Taylor lower bounds around $(p_u^{(j)},q_u^{(j)})$, $(p_t^{(j)},q_t^{(j)})$, and $(\mathbf{p}_r^{(j)},\mathbf{q}_r^{(j)})$, respectively, at the $j$th SCA iteration, as follows:
\begin{subequations}
\begin{align}
&\big(p_u^{(j)}\big)^2
+2p_u^{(j)}(p_u-p_u^{(j)})
+\big(q_u^{(j)}\big)^2
+2q_u^{(j)}(q_u-q_u^{(j)})
\ge \frac{\rho_u}{\gamma_u}, \label{eq:P3sca_user}\\
&\big(p_t^{(j)}\big)^2
+2p_t^{(j)}(p_t-p_t^{(j)})
+\big(q_t^{(j)}\big)^2
+2q_t^{(j)}(q_t-q_t^{(j)})
\ge \frac{\rho_t}{\gamma_r}, \label{eq:P3sca_target}\\
&2\big(\mathbf{p}_r^{(j)}\big)^T\mathbf{p}_r
-\left\|\mathbf{p}_r^{(j)}\right\|^2
+2\big(\mathbf{q}_r^{(j)}\big)^T\mathbf{q}_r
-\left\|\mathbf{q}_r^{(j)}\right\|^2
\ge \frac{\zeta_t}{|\beta|^2}.
\label{eq:P3sca_receive}
\end{align}
\end{subequations}

Next, we address the sparsity constraints \eqref{eq:P1_c2} and \eqref{eq:P1_c3} and the binary constraints. To obtain a tractable formulation, we employ reweighted $\ell_1$ minimization \cite{EnhancingSparsity,Yadav2021NOMA}, relax the binary constraints in \eqref{eq:P1_c5} and \eqref{eq:P1_c6}, and introduce sparsity-promoting penalties into the objective. Accordingly, at the $j$th SCA iteration, problem~\eqref{prob:P3} is relaxed to obtain the TPA and RPA activation variables as follows:
\begin{subequations}\label{prob:P3_relaxed}
    \begin{align}
        \max_{\substack{ \mathbf{a},\,\mathbf{b}, \rho_u,\\ \rho_t,\,\zeta_t}} 
        & \log_2(1+\rho_u)
        -\mu_{\text{T}} \sum_{m=1}^{M}\sum_{k=1}^{K_{\text{T}}}
        \lambda_{m,k}^{(j)} a_{m,k} -\mu_{\text{R}} \sum_{n=1}^{N}\sum_{l=1}^{K_{\text{R}}}
        \lambda_{n,l}^{(j)} b_{n,l} \\
        \text{s.t.}\quad
        & \|\mathbf{a}_m\|_1 \leq N_{\mathrm{act},\text{T}}, \quad \forall m\in\mathcal{M}, \label{eq:P3_relaxed_c1}\\
        & \|\mathbf{b}_n\|_1 \leq N_{\mathrm{act},\text{R}}, \quad \forall n\in\mathcal{N}, \label{eq:P3_relaxed_c2}\\        
        & 0 \le a_{m,k} \le 1,\quad \forall m,k, \label{eq:P3_relaxed_box_T}\\
        & 0 \le b_{n,l} \le 1,\quad \forall n,l, \label{eq:P3_relaxed_box_R}\\
        & \eqref{eq:P3_bilinear_sca},\ \eqref{eq:P3_pu_qu},\ \eqref{eq:P3_pt_qt},\ \eqref{eq:P3_pr_qr},\ \eqref{eq:P3sca_user},\ \eqref{eq:P3sca_target},\ \eqref{eq:P3sca_receive},
    \end{align}
\end{subequations}
where $\mu_{\text{T}}$ and $\mu_{\text{R}}$ control the sparsity-promoting penalties for the transmit and receive activation variables, respectively, while $\lambda_{m,k}^{(j)}$ and $\lambda_{n,l}^{(j)}$ are the penalty weights updated as $\lambda_{m,k}^{(j)}=\frac{1}{a_{m,k}^{(j-1)}+\epsilon}$ and $\lambda_{n,l}^{(j)}=\frac{1}{b_{n,l}^{(j-1)}+\epsilon}$ with a small value of $\epsilon$. 

Problem~\eqref{prob:P3_relaxed} is convex and can be solved efficiently using standard convex optimization solvers. The overall AO algorithm for solving the proposed joint discrete pinching-antenna activation and beamforming optimization (JDPABO) problem is summarized in Algorithm~\ref{alg:ao}.
{\footnotesize
\begin{algorithm}[t]
\caption{Proposed JDPABO Algorithm}
\label{alg:ao}
    \begin{algorithmic}[1]
        \REQUIRE Initial feasible $\mathbf{w}^{(0)}$, $\mathbf{a}^{(0)}$, and $\mathbf{b}^{(0)}$; penalty parameters $\mu_{\text{T}}$ and $\mu_{\text{R}}$; small constant $\epsilon>0$; tolerance $\varepsilon_{\mathrm{tol}}$.
        \STATE Set AO iteration index $i=0$. 
        \REPEAT
        \STATE For fixed $\mathbf{A}^{(i)}$ and $\mathbf{B}^{(i)}$, solve \eqref{prob:P2_sca} iteratively until convergence to obtain $\mathbf{w}^{(i+1)}$.
        \STATE For fixed $\mathbf{w}^{(i+1)}$, solve \eqref{prob:P3_relaxed} iteratively until convergence to obtain $\mathbf{a}^{(i+1)}$ and $\mathbf{b}^{(i+1)}$.
        \STATE Update $\mathbf{A}^{(i+1)}=\mathrm{diag}(\mathbf{a}^{(i+1)})$ and $\mathbf{B}^{(i+1)}=\mathrm{diag}(\mathbf{b}^{(i+1)})$.
        \STATE Set $i=i+1$.
        \UNTIL{ the objective improvement falls below $\varepsilon_{\mathrm{tol}}$.}
        \ENSURE $\mathbf{w}^{(i)}$, $\mathbf{a}^{(i)}$, and $\mathbf{b}^{(i)}$.
    \end{algorithmic}
\end{algorithm}
}
The computational complexity of the proposed algorithm is mainly determined by solving the beamforming and antenna-activation subproblems. Using the interior-point method, the worst-case computational complexity per outer JDPABO iteration is $\mathcal{O}\!\left(\left[J_w M^3 + J_{ab}\left(MK_{\text{T}} + NK_{\text{R}}\right)^{3}\right]\log\left(\frac{1}{\delta}\right)\right),$ where $J_w$ and $J_{ab}$ are the numbers of SCA iterations required for the beamforming and antenna-activation subproblems, respectively, and $\delta$ denotes the solution accuracy.

\section{Simulation Results}
\begin{figure*}
\vspace{-8mm}
\begin{minipage}[t]{0.32\textwidth}
  \includegraphics[width=\linewidth]{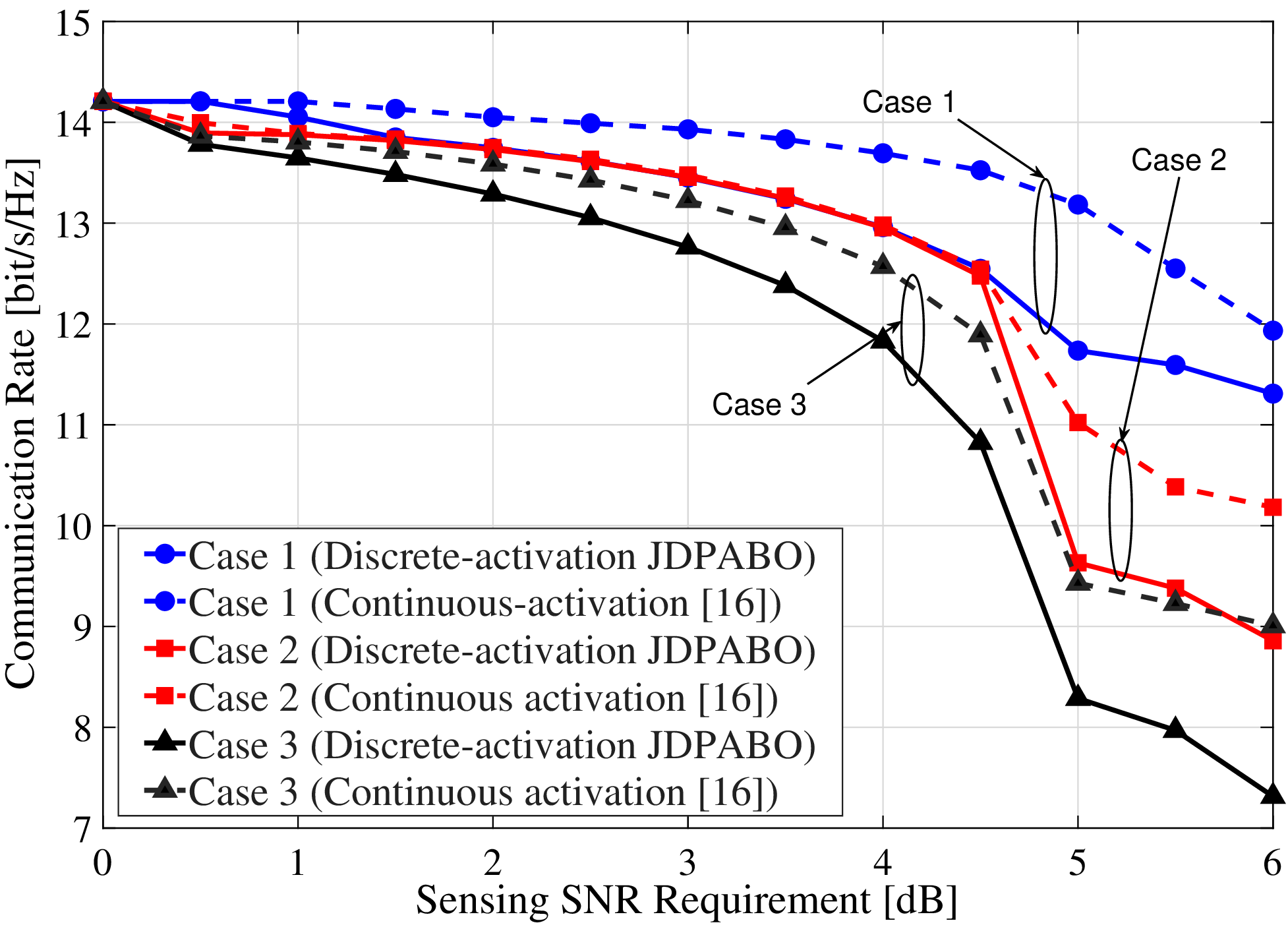}
  \caption{Communication rate versus sensing SNR requirement.}
  \label{fig:rate_vs_snr_cases}
\end{minipage}%
\hfill 
\begin{minipage}[t]{0.32\textwidth}
  \includegraphics[width=\linewidth]{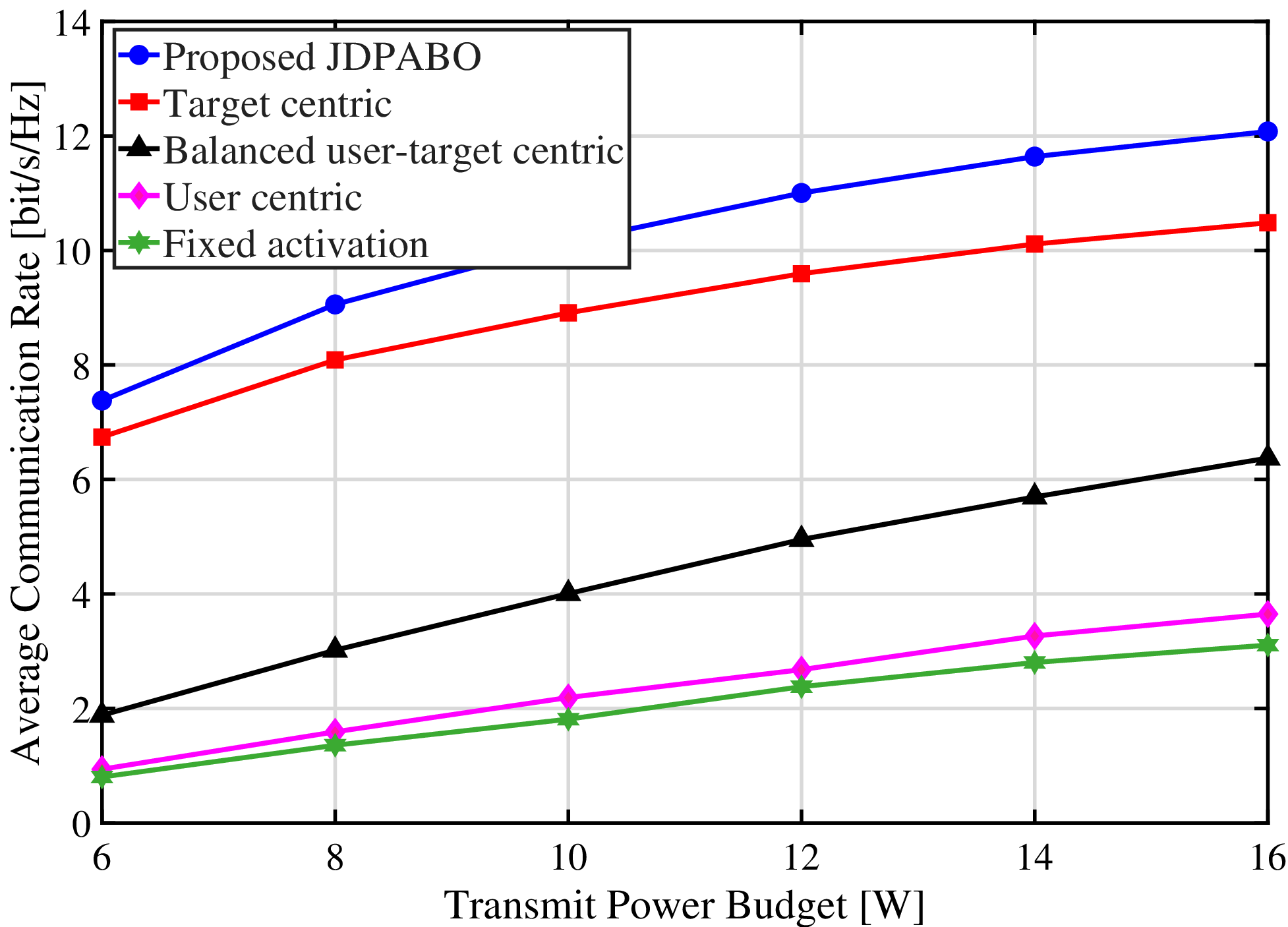}
  \caption{Communication rate versus transmit power with $\gamma_{s,\min}=4$ dB.}
  \label{fig:rate_vs_transmit_power}
\end{minipage}%
\hfill
\begin{minipage}[t]{0.32\textwidth}
  \includegraphics[width=\linewidth]{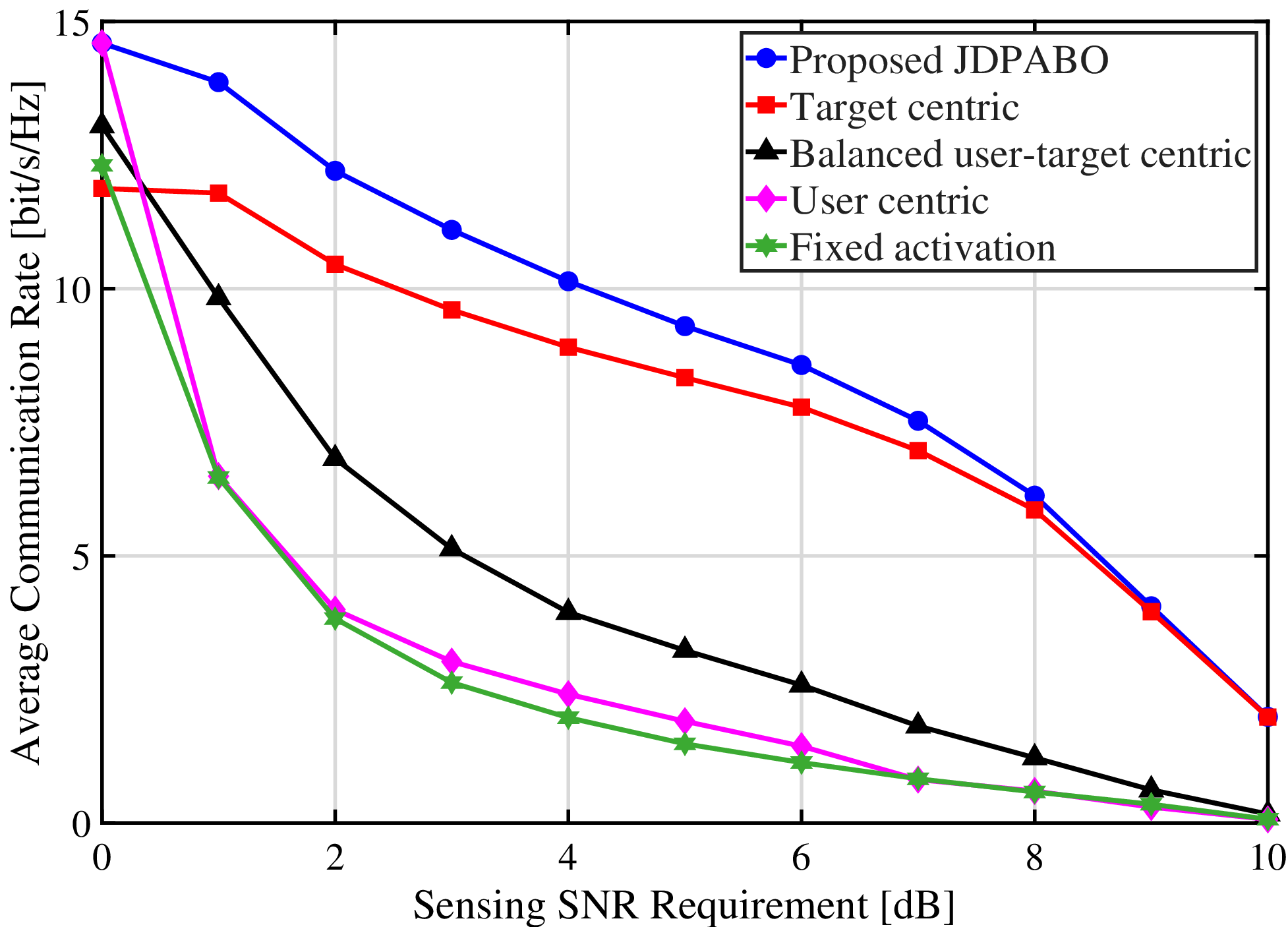}
  \caption{Communication rate versus sensing SNR requirement under random user and target locations.}
      \label{fig:rate_vs_snr_cases_2}
\end{minipage}%
\end{figure*}

In this section, we evaluate the performance of the proposed JDPABO algorithm. Unless otherwise stated, the simulation parameters are set as follows: $L=40~\mathrm{m}$, $W=20~\mathrm{m}$, $H=3~\mathrm{m}$, $K_{\text{T}}=K_{\text{R}}=81$, $c=3\times10^{8}~\mathrm{m/s}$, $n_{\mathrm{eff}}=1.4$, $M=N=4$, $N_{\mathrm{act},\text{T}}=N_{\mathrm{act},\text{R}}=1$, $\lambda=5~\mathrm{cm}$, $|\beta|=1$, $P_{\max}=10~\mathrm{W}$, $\gamma_{s,\min}=4$ dB, $\sigma_u^2=-60~\mathrm{dBm}$, and $\sigma_r^2=-80~\mathrm{dBm}$. To evaluate the impact of the relative user and target locations, we considered three representative cases. In Case 1, the user and target coordinates are set to $\boldsymbol{\psi}_u=[L/10,2W/5,0]^T$ and $\boldsymbol{\psi}_t=[-L/10,3W/5,0]^T$. In Case 2, they are set to $\boldsymbol{\psi}_u=[L/5,2W/5,0]^T$ and $\boldsymbol{\psi}_t=[-L/5,3W/5,0]^T$. In Case 3, they are set to $\boldsymbol{\psi}_u=[3L/10,2W/5,0]^T$ and $\boldsymbol{\psi}_t=[-3L/10,3W/5,0]^T$. For the comparison in Fig.~\ref{fig:rate_vs_snr_cases}, we set $M=2$ to match the number of transmit
waveguides considered in \cite{mao2025multiWaveguidePASSISAC}.

\begin{figure}[t]
    \centering
    \includegraphics[width=0.99\linewidth]{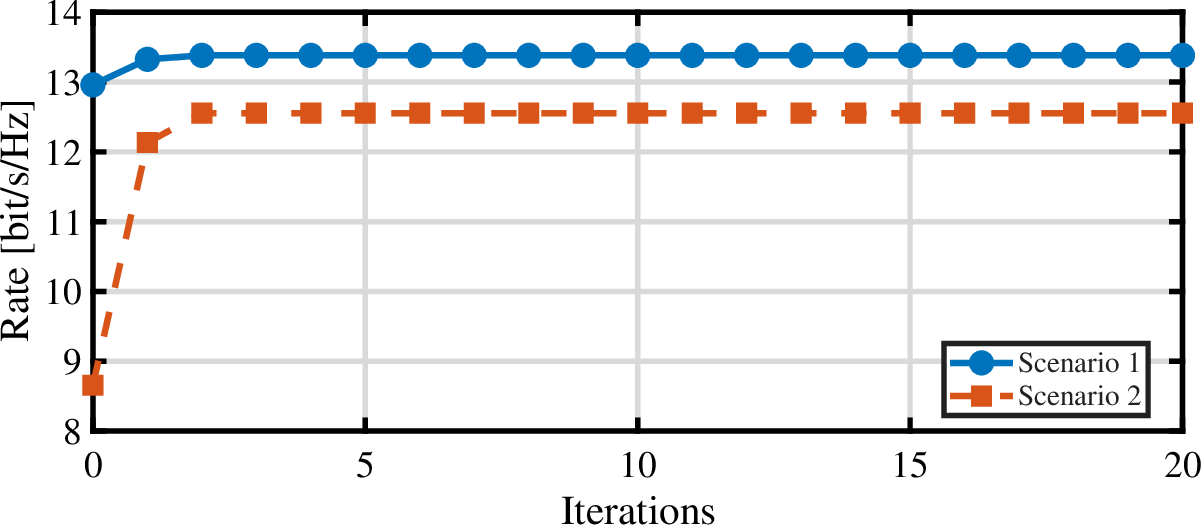}
    \caption{Convergence behavior of JDPABO at $\gamma_{s,\min}=4$ dB.}
    \label{fig:convergence_plot}
\end{figure}

 Fig.~\ref{fig:convergence_plot} plots the communication rate versus the number of AO iterations for the proposed JDPABO algorithm at $\gamma_{s,\min}=4$ dB under two different scenarios. It can be observed that the JDPABO algorithm reaches convergence within a few iterations in both scenarios.

Fig.~\ref{fig:rate_vs_snr_cases} plots and compares the communication rate of the proposed JDPABO algorithm with the continuous-location multi-waveguide PASS-ISAC benchmark \cite{mao2025multiWaveguidePASSISAC} under different sensing SNR requirements for the three representative cases. JDPABO closely approaches the benchmark at low sensing SNR requirements. However, the continuous-location benchmark achieves a higher rate owing to its greater flexibility in antenna placement, whereas discrete activation is restricted to predefined candidate locations. This performance gap becomes more pronounced at higher sensing SNR requirements, particularly in Cases 2 and 3, where the user and target are farther apart. Moreover, the abrupt rate reductions at high $\gamma_{s,\min}$ indicate that the activated TPAs must shift from user-favorable toward target-favorable locations to satisfy the sensing constraint, resulting in a larger communication-rate penalty.
  
To further evaluate the proposed system, we considered four baseline schemes. \textit{Baseline-1 (Fixed-Antenna):} active TPAs are selected near the center of the serving region, without adapting to the user or target locations. \textit{Baseline-2 (User Centric):} active TPAs are selected near the user to prioritize communication performance. \textit{Baseline-3 (Target Centric):} active TPAs are selected near the target to prioritize sensing performance. \textit{Baseline-4 (Balanced User-Target Centric):} active TPAs are selected near the midpoint between the user and target to balance communication and sensing performance.

Fig.~\ref{fig:rate_vs_transmit_power} compares the communication rate of JDPABO with the baseline schemes as the transmit power increases. JDPABO consistently achieves the highest rate by jointly optimizing transmit beamforming and TPA/RPA activation based on the user and sensing channels. The target-centric scheme performs closest to JDPABO, as activating TPAs near the target improves illumination and facilitates satisfaction of the sensing SNR constraint. The balanced user-target centric scheme provides a less favorable communication-sensing balance, while the user-centric scheme improves the user link at the expense of target illumination. The fixed activation scheme performs worst due to its inability to adapt the antenna activation to the user-target geometry.

Fig.~\ref{fig:rate_vs_snr_cases_2} shows the communication rate versus the sensing SNR requirement for random user and target locations. As $\gamma_{s,\min}$ increases, the rate decreases for all schemes, highlighting the communication-sensing trade-off. At low sensing requirements, the user-centric scheme remains competitive by strengthening the communication link. As the sensing requirement increases, the target-centric scheme approaches the proposed scheme by providing stronger target illumination. In contrast, the midpoint-based and fixed schemes exhibit larger performance gaps because their activation patterns are less adaptive to the user-target geometry.

\section{Conclusion}
In this paper, we investigated a discrete-activation multi-waveguide PASS-ISAC architecture, where TPAs/RPAs are selected from pre-installed candidate locations. We developed a low-complexity JDPABO algorithm to jointly optimize the waveguide beamforming weights and sparse TPA/RPA activation patterns for communication-rate maximization under transmit-power, antenna-selection, and sensing SNR constraints. Simulation results demonstrated that JDPABO outperformed fixed activation schemes and approached the continuous-placement benchmark at low sensing SNR requirements.

\bibliographystyle{IEEEtran}
\bibliography{PASS_ref.bib}

@article{wen2024iscc,
  author  = {D. Wen and Y. Zhou and X. Li and Y. Shi and K. Huang and K. B. Letaief},
  title   = {A Survey on Integrated Sensing, Communication, and Computation},
  journal = {IEEE Commun. Surveys Tuts.},
  volume  = {27},
  number  = {5},
  pages   = {3058--3098},
  month   = oct,
  year    = {2025}
}

@article{liu2025passOutlook,
  author  = {Y. Liu and Z. Wang and X. Mu and C. Ouyang and X. Xu and Z. Ding},
  title   = {Pinching-Antenna Systems: Architecture Designs, Opportunities, and Outlook},
  journal = {IEEE Commun. Mag.},
  volume  = {64},
  number  = {1},
  pages   = {190--196},
  month   = jan,
  year    = {2026}
}

@article{pan2020ris6g,
  author  = {C. Pan and others},
  title   = {Reconfigurable Intelligent Surfaces for 6{G} Systems: Principles, Applications, and Research Directions},
  journal = {IEEE Commun. Mag.},
  volume  = {59},
  number  = {6},
  pages   = {14--20},
  month   = jun,
  year    = {2021}
}

@article{wu2024fas6g,
  author  = {T. Wu and others},
  title   = {Fluid Antenna Systems Enabling 6{G}: Principles, Applications, and Research Directions},
  journal = {IEEE Wireless Commun.},
  volume  = {33},
  number  = {4},
  pages   = {100--108},
  month   = aug,
  year    = {2026}
}

@article{zhu2023ma,
  author  = {L. Zhu and W. Ma and B. Ning and R. Zhang},
  title   = {Movable-Antenna Enhanced Multiuser Communication via Antenna Position Optimization},
  journal = {IEEE Trans. Wireless Commun.},
  volume  = {23},
  number  = {7},
  pages   = {7214--7229},
  month   = jul,
  year    = {2024}
}

@article{yang2025pinchingPrinciples,
  author  = {Z. Yang and others},
  title   = {Pinching Antennas: Principles, Applications and Challenges},
  journal = {IEEE Wireless Commun.},
  volume  = {33},
  number  = {2},
  pages   = {175--184},
  month   = apr,
  year    = {2026}
}

@article{suzuki2022pinching,
  author  = {A. Fukuda and H. Yamamoto and H. Okazaki and Y. Suzuki and K. Kawai},
  title   = {Pinching Antenna: Using a Dielectric Waveguide as an Antenna},
  journal = {NTT DOCOMO Tech. J.},
  volume  = {23},
  number  = {3},
  pages   = {5--12},
  month   = jan,
  year    = {2022}
}

@article{ouyang2025passModeling,
  author  = {Z. Wang and C. Ouyang and X. Mu and Y. Liu and Z. Ding},
  title   = {Modeling and Beamforming Optimization for Pinching-Antenna Systems},
  journal = {IEEE Trans. Commun.},
  volume  = {73},
  number  = {12},
  pages   = {13904--13919},
  month   = dec,
  year    = {2025}
}

@article{ding2024pinchingPerspective,
  author  = {Z. Ding and R. Schober and H. V. Poor},
  title   = {Flexible-Antenna Systems: A Pinching-Antenna Perspective},
  journal = {IEEE Trans. Commun.},
  volume  = {73},
  number  = {10},
  pages   = {9236--9253},
  month   = oct,
  year    = {2025}
}

@article{zhang2025passISAC,
  author  = {Z. Zhang and Z. Wang and X. Mu and B. He and J. Chen and Y. Liu},
  title   = {Integrated Sensing and Communications for Pinching-Antenna Systems ({PASS})},
  journal = {IEEE Commun. Lett.},
  volume  = {29},
  number  = {12},
  pages   = {2929--2933},
  month   = dec,
  year    = {2025}
}

@article{ding2025crlbPASSISAC,
  author  = {Z. Ding},
  title   = {Pinching-Antenna Assisted {ISAC}: A {CRLB} Perspective},
  journal = {npj Wireless Technol.},
  volume  = {1},
  number  = {4},
  pages   = {1--7},
  month   = {Nov.},
  year    = {2025}
}

@article{mao2025multiWaveguidePASSISAC,
  author  = {W. Mao and Y. Lu and Y. Xu and B. Ai and O. A. Dobre and D. Niyato},
  title   = {Multi-Waveguide Pinching Antennas for {ISAC}},
  journal = {IEEE Trans. Wireless Commun.},
  volume  = {25},
  pages   = {5846--5858},
  month   = {Oct.},
  year    = {2026}
}

@article{qin2025passISACPositionPower,
  author  = {Y. Qin and Y. Fu and H. Zhang},
  title   = {Joint Antenna Position and Transmit Power Optimization for Pinching Antenna-Assisted {ISAC} Systems},
  journal = {IEEE Wireless Commun. Lett.},
  volume  = {14},
  number  = {11},
  pages   = {3535--3539},
  month   = nov,
  year    = {2025}
}

@article{xu2025passPowerRadiation,
  author  = {X. Xu and X. Mu and Z. Wang and Y. Liu and A. Nallanathan},
  title   = {Pinching-Antenna Systems ({PASS}): Power Radiation Model and Optimal Beamforming Design},
  journal = {IEEE Trans. Commun.},
  volume  = {74},
  pages   = {2160--2175},
  month   = nov,
  year    = {2026}
}

@article{EnhancingSparsity,
  author  = {E. J. Cand{\`e}s and M. B. Wakin and S. P. Boyd},
  title   = {Enhancing Sparsity by Reweighted {$\ell_1$} Minimization},
  journal = {J. Fourier Anal. Appl.},
  volume  = {14},
  number  = {5--6},
  pages   = {877--905},
  month   = {Oct.},
  year    = {2008}
}

@article{zhang2022beamfocusing,
  author  = {H. Zhang and others},
  title   = {Beam Focusing for Near-Field Multiuser {MIMO} Communications},
  journal = {IEEE Trans. Wireless Commun.},
  volume  = {21},
  number  = {9},
  pages   = {7476--7490},
  month   = sep,
  year    = {2022}
}

@article{Marks1978,
  author  = {B. R. Marks and G. P. Wright},
  title   = {A General Inner Approximation Algorithm for Nonconvex Mathematical Programs},
  journal = {Oper. Res.},
  volume  = {26},
  number  = {4},
  pages   = {681--683},
  month   = {Jul.-Aug.},
  year    = {1978}
}

@article{Yadav2021NOMA,
  author  = {A. Yadav and others},
  title   = {On Performance Comparison of Multi-Antenna {HD-NOMA}, {SCMA}, and {PD-NOMA} Schemes},
  journal = {IEEE Wireless Commun. Lett.},
  volume  = {10},
  number  = {4},
  pages   = {715--719},
  month   = apr,
  year    = {2021}
}

@article{Wijewardhana2026WiPASS,
  author  = {K. R. Wijewardhana and A. Yadav and M. Zeng and M. Elsayed and O. A. Dobre and Z. Ding},
  title   = {Wireless-Fed Pinching-Antenna Systems ({Wi-PASS}) for {NextG} Wireless Networks},
  journal = {IEEE Commun. Mag.},
  pages   = {1--7},
  year    = {2026},
  note    = {{E}arly Access}
}

@article{Zeng2026PASSRA,
  author  = {M. Zeng and others},
  title   = {Resource Allocation for Pinching-Antenna Systems: State-of-the-Art, Key Techniques and Open Issues},
  journal = {IEEE Wireless Commun.},
  pages   = {1--7},
  year    = {2026},
  note    = {{E}arly Access}
}

@article{Xu2026GeneralizedPASS,
  author  = {Y. Xu and others},
  title   = {Generalized Pinching-Antenna Systems: A Tutorial on Principles, Design Strategies, and Future Directions},
  journal = {IEEE Commun. Surveys Tuts.},
  volume  = {28},
  pages   = {5872--5908},
  year    = {2026}
}

@article{Yang2026PinchingAntennas,
  author  = {Z. Yang and others},
  title   = {Pinching Antennas: Principles, Applications and Challenges},
  journal = {IEEE Wireless Commun.},
  volume  = {33},
  number  = {2},
  month   = {Apr.},
  pages   = {175--184},
  year    = {2026}
}

@article{Wang2025AntennaActivation,
  author  = {K. Wang and Z. Ding and R. Schober},
  title   = {Antenna Activation for {NOMA} Assisted Pinching-Antenna Systems},
  journal = {IEEE Wireless Commun. Lett.},
  volume  = {14},
  number  = {5},
  pages   = {1526--1530},
  month   = may,
  year    = {2025},
  doi     = {10.1109/LWC.2025.3548280}
}

@article{Zhao2026WaveguideDivision,
  author  = {J. Zhao and X. Mu and K. Cai and Y. Zhu and Y. Liu},
  title   = {Waveguide Division Multiple Access for Pinching-Antenna Systems ({PASS})},
  journal = {IEEE Trans. Wireless Commun.},
  volume  = {25},
  pages   = {13761--13775},
  year    = {2026}
}

\end{document}